\documentclass[conference]{IEEEtran}
\IEEEoverridecommandlockouts
\usepackage[pdftex]{graphicx}
\usepackage[ruled]{algorithm2e}
\usepackage{algorithmic}
\usepackage[small]{caption}
\usepackage{float}
\usepackage{xspace}
\usepackage{fancyhdr}
\usepackage{amssymb,amsmath}
\usepackage{subfig}
\usepackage{epstopdf}
\usepackage{bbding}
\usepackage{cite} 
\usepackage{color}
\begin{document}

\title{\huge Rate Splitting Multiple Access for Cognitive Radio \\GEO-LEO Co-Existing
Satellite Networks}
\author{Wali Ullah Khan$^1$, Zain Ali$^2$, Eva Lagunas$^1$, Symeon Chatzinotas$^1$, Bj\"orn Ottersten$^1$ \\$^1$Interdisciplinary Centre for Security, Reliability and Trust (SnT), University of Luxembourg\\
$^2$Department of Electrical and Computer Engineering, University of California, Santa Cruz, USA\\
\{waliullah.khan, eva.lagunas, symeon.chatzinotas, bjorn.ottersten\}@uni.lu\\
zainalihanan1@gmail.com
\thanks{This work has been supported by the Luxembourg National Research Fund (FNR) under the project MegaLEO (C20/IS/14767486).}

}%

\maketitle

\begin{abstract}
Low Earth orbit (LEO) satellite communication has drawn particular attention recently due to its high data rate services and low round-trip latency. It is low-cost to launch and can provide global coverage. However, the spectrum scarcity might be one of the critical challenges in the growth of LEO satellites, impacting severe restrictions on the development of ground-space integrated networks. To address this issue, we propose rate splitting multiple access (RSMA) for cognitive radio (CR) enabled nongeostationary orbit (GEO)-LEO coexisting satellite network. In particular, this work aims to maximize the system's sum rate by simultaneously optimizing the power allocation and subcarrier beam assignment of LEO satellite communication while restricting the interference temperature to GEO satellite users. The problem of sum rate maximization is formulated as non-convex and a Global optimal solution is challenging to obtain. Therefore, we first employ the successive convex approximation technique to reduce the complexity and make the problem more tractable. Then for the power allocation, we exploit Karush–Kuhn–Tucker (KKT) condition and adopt an efficient algorithm based on the greedy approach for subcarrier beam assignment. We also propose two suboptimal schemes with fixed power allocation and random subcarrier beam assignment as the benchmark. Results demonstrate the benefits of the proposed scheme compared to the benchmark schemes.  
\end{abstract}

\begin{IEEEkeywords}
GEO satellite, LEO satellite, cognitive radio, rate splitting multiple access, spectral efficiency optimization.
\end{IEEEkeywords}

\section{Introduction}
Satellite communication has recently gained significant attention in industry and academia due to its capability to provide global coverage and support a wide range of services \cite{9460776}. Three existing satellite communication types are geosynchronous equatorial orbit (GEO) satellite, medium Earth orbit (MEO) satellite, and low Earth orbit (LEO) satellite, respectively. Due to the low orbit profile, the LEO satellite has the ability to provide high-speed data, and low round-trip latency \cite{9512414}. Moreover, its launching cost is comparatively lower than GEO and MEO, making it more likely to achieve global coverage. However, the increasing demand for different services would require many satellites in different orbits, which might be challenging using limited spectrum resources \cite{9222141}. As a result, it can seriously affect the future developments in coexisted ground-space communication networks. One of the potential approaches which might be helpful in this situation is the efficient spectrum sharing across different orbits using advanced multiple access techniques \cite{9210567}.

Cognitive radio (CR) and rate splitting multiple access (RSMA) have emerged as promising technologies for providing high spectral efficiency and have the potential to ease the above situations \cite{8957541,8357810}. These technologies can simultaneously accommodate multiple users over the same spectrum and time resources, which significantly enhances the spectral efficiency of the system. More specifically, in CR, the licensed primary and unlicensed secondary networks communicate over the same spectrum such that the secondary network would not cause harmful interference to the primary network \cite{6845054}. On the other side, the fundamental concept of RSMA is to accommodate multiple users over the same spectrum and time resources. According to RSMA protocol, the signal transmitted to the users is divided into two signals, i.e., the common part and the private part \cite{9461768}. The common parts of the signals can be combined into a single common signal first and then encoded with a public shared code-book. In contrast, the private parts of the signals can be independently encoded to specific intended users. Each user first decodes the common part of the signal using the shared code-book. Then each user reconstructs its original signal from the part of its common and intended private signals using the successive interference cancellation (SIC) technique. 

Recently, researchers have proposed RSMA for satellite communications. For example, Yin {\em et al.} \cite{9145200,9257433} have considered RSMA for GEO satellite communication and solved the max-min fairness problem using Weighted Minimum-Mean Square Error approach. Lin {\em et al.} \cite{9324793} have considered RSMA for GEO unmanned aerial vehicle (UAV) integrated networks. They investigate a sum rate maximization problem using sequential convex approximation and the first-order Taylor expansion approaches. Moreover, the works in \cite{9473795,9684855} have also investigated the max-min data rate  problem in GEO satellite communications. They consider RSMA beamforming and two-stage precoding schemes with imperfect channel state information (CSI). It can be observed that the works in \cite{9145200,9257433,9324793,9473795,9684855} have considered RSMA only in GEO satellite communication, and they do not consider cognitive radio. To the best of our knowledge, the work of spectral efficiency optimization that considers cognitive radio GEO-LEO coexisting satellite network with the RSMA technique has not yet been investigated in the literature. To bridge this open gap, we consider a cognitive radio GEO-LEO coexisting satellite communication network RSMA.

This work aims to maximize the sum rate of cognitive radio GEO-LEO coexisting satellite communications. We simultaneously optimize the power budgets over different beams of LEO satellite, power allocation coefficients for ground users over each beam based on RSMA protocol, and subcarrier beam assignment to the ground users. This framework is subjected to the interference temperature at GEO satellite users from LEO satellite and the quality of services of LEO users. The sum rate maximization problem is non-convex and hard to obtain the Global optimal solution. To reduce the complexity and make the problem tractable, we first adopt the successive convex approximation technique, where a properly chosen surrogate can efficiently replace the original non-convex function. Then we apply Karush–Kuhn–Tucker (KKT) conditions for optimal power allocation and design an algorithm for efficient subcarrier beam assignment based on the greedy approach. We also propose two suboptimal schemes with a fixed power budget at each beam and random subcarrier beam assignment as the benchmark. The remaining of this work is structured as follows. Section II discusses system model and problem formulation. Sections III and IV provide proposed solution and simulation results while Section V concludes this work.
\begin{figure}[!t]
\centering
\includegraphics[width=0.35\textwidth]{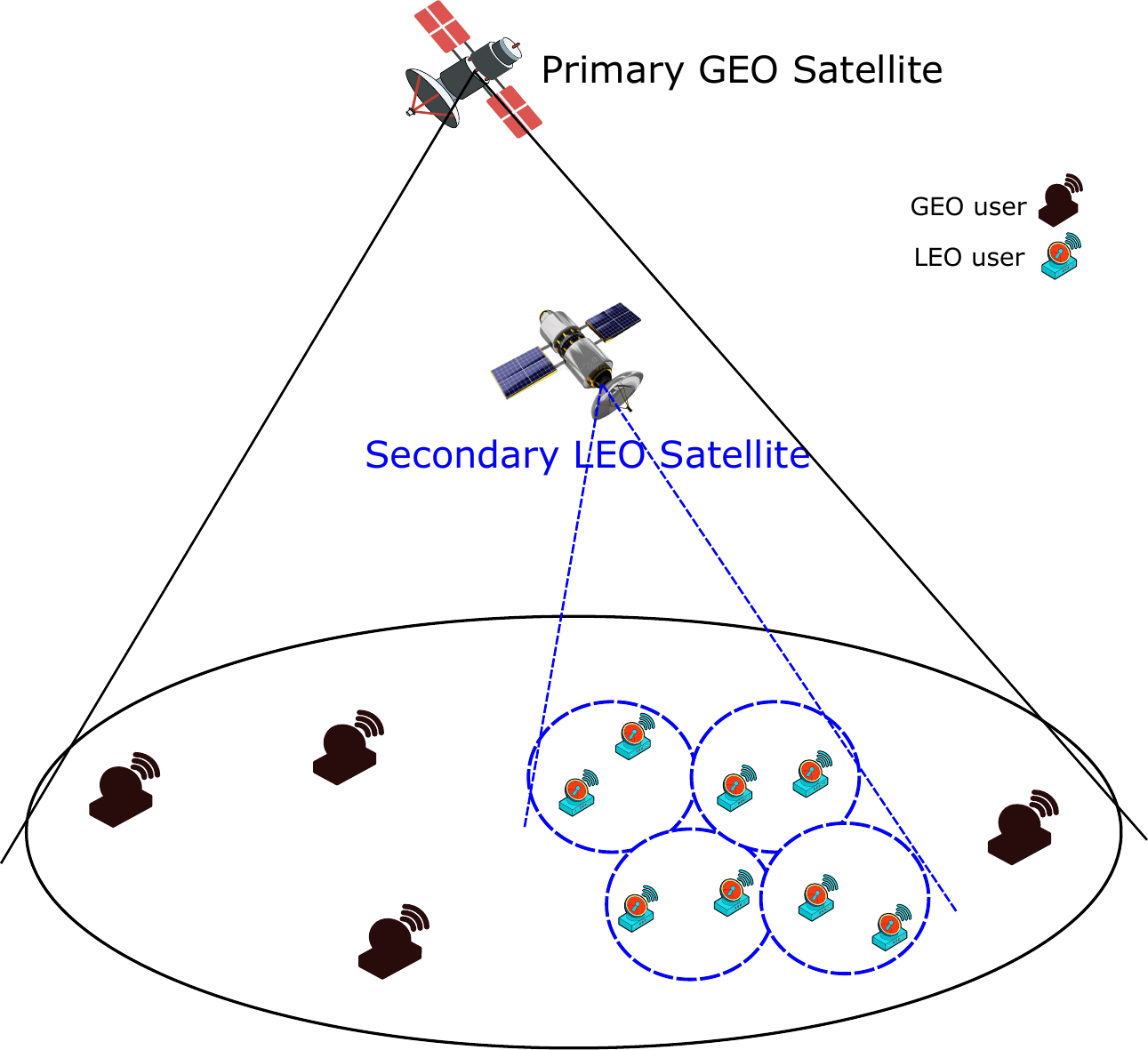}
\caption{System Model}
\label{blocky}
\end{figure}
\section{System Model and Problem Formulation}
As shown in Fig. \ref{blocky}, we consider a multi-carrier CR-inspired GEO-LEO co-existing satellite network such that a primary GEO satellite shares $K$ subcarrier with a secondary LEO satellite. Both GEO satellite and LEO satellite follow multibeam technology, where GEO users and LEO users are randomly located in a beam coverage of the GEO satellite. In the considered scenario, GEO satellite accommodates users through the OMA technique, while LEO adopts RSMA to serve its associated users. If $U$ denotes the set of total LEO users and $U_m$ is the subsect of users associated with $m$ bean through $k$ subcarrier (i.e., $U_m\in U$), then the common signal of $U_m$ is denoted as $s_{m,0,k}$ while the private message of $u$ user is stated as $s_{m,u,k}$, where $u\in U_m$. Moreover, the common part of all users signals associated with $m$ beam can be combined into a single common signal. The single common signal and $U_m$ private signals can be then independently encoded into streams $s_{m,0,k}, s_{m,1,k}, s_{m,2,k}, \dots,s_{m,U,k}$, where $s_{m,0,k}$ and $s_{m,u,k}$ represent the encoded common and private data symbols. The overall transmitted signal of $m$ beam to $U_m$ users over $k$ subcarrier can be written as
\begin{align}
s_{m,k}=\sqrt{\eta_{m,0,k}p_{m,k}}s_{m,0,k}+\sum\limits_{u=1}^{U_m}\sqrt{\eta_{m,u,k}p_{m,k}}s_{m,u,k},
\end{align}
where $p_{m,k}$ denotes the transmit power of $m$ beam over $k$ channel. $\eta_{m,0,k}$ and $\eta_{m,u,k}$ are the power allocation coefficients for the common message $s_{m,0,k}$ and private message $s_{m,u,k}$ over $k$ subcarrier. We model the channel gain from $m$ LEO beam to $u$ user over $k$ subcarrier as $h_{m,u,k}=G_T G_R\div L_{m,u,k}$. Where $G_T$ and $G_R$ are the gains of transmit and received antennas while $L_{m,u,k}=(4\pi D\div(c/f))^2$ denotes the free space propagation loss. The signal that user $u$ received from $m$ LEO beam over $k$ subcarrier can be expressed as
\begin{align}
y_{m,u,k}&=\sqrt{h_{m,u,k}p_{m,k}\eta_{m,0,k}}s_{m,0,k}+\omega_{j,k}\nonumber\\&+\sum\limits_{i=1}^{U_m}\sqrt{h_{m,u,k}\eta_{m,i,k}p_{m,k}}s_{m,i,k}\nonumber\\&+\sqrt{g^{m'}_{m,u,k}q_{m',u',k}}e_{m',u',k},\label{3}
\end{align}  
where $g^{m'}_{m,u,k}$ is the channel gain from the GEO satellite to the $u$ user over $k$ subcarrier and $q_{m',u',k}$ its associated transmit power. Further, $e_{m',u',k}$ denotes the transmitted signal of GEO satellite to its $u'$ user over $k$ subcarrier, and $\omega_{m,u,k}$ is the additive noise at $u$ user. To ensure that the common signal can be successfully
decoded by all users associated with $m$ beam of LEO satellite over $k$ subcarrier, the achievable data rate of the common message can be stated as
\begin{align}
R_{m,c,k}= \underset{u\in U}{\min} W\log_2\bigg(1+\frac{h_{m,u,k}\eta_{m,0,k}p_{m,k}}{g^{m'}_{m,u,k}q_{m',u',k}+I_{0}+\sigma^2}\bigg),
\end{align} 
where $I_{0}=h_{m,u,k}\sum_{j=1}^{U} x_{m,j,k}\eta_{m,j,k}p_{m,k}$ is the interference of RSMA users during decoding of common signal, $W$ is the bandwidth available at $m$ LEO beam and $\sigma^2$ denotes the noise variance. Moreover, $x_{m,j,k}$ is the binary variable for subcarrier beam assignment. Next we define the binary variable for subcarrier beam assignment such that
\begin{align*}
x_{m,u,k}=\left\{ 
  \begin{array}{ c l }
    1 & \quad \textrm{if $u$ user is assigned to beam $m$ over $k$} \\
    &\quad\textrm{subcarrier},\\
    0                 & \quad \textrm{otherwise}.
  \end{array}
\right.
\end{align*}

Since $R_{m,c,k}$ is the shared signal rate between users on $k$ subcarrier such that $ C_{m,u,k}$ denotes the portion of $u$'s user data rate, where $\sum_{u=1}^{U} C_{m,u,k}\leq  R_{m,c,k}$. After successfully decoding the common signal $s_{m,0,k}$, each user also decodes its private signal, the achievable data rate of $u$ user to decode its private signal $s_{m,u,k}$ can be written as
\begin{align}
R_{m,u,k}= W\log_2\bigg(1+\frac{h_{m,u,k}\eta_{m,u,k}p_{m,k}}{g^{m'}_{m,u,k}q_{m',u',k}+I_{u}+\sigma^2}\bigg),
\end{align}
where $I_{u}=h_{m,u,k}\sum_{j=1,j\neq u}^{U}x_{m,j,k}\eta_{m,j,k}p_{m,k}$ denotes the RSMA interference during decoding the $u$ user private signal.
Given the achievable data rate of common and private signals, the total achievable data rate of $u$ user from $m$ LEO beam over $k$ subcarrier can be written as $ R_{tot}=  C_{m,u,k}+ R_{m,u,k}$. 

To protect the quality of services of GEO satellite user over $k$ subcarrier from the interference of $m$ LEO beam, our optimization framework restricts interference temperature as
\begin{align}
f^{m}_{m',u',k} p_{m,k}\leq I_{th}, \forall m,k
\end{align}
where $I_{th}$ is the maximum interference temperature threshold to GEO user over $k$ subcarrier.

Given the proposed system model, we seek to optimize the spectral efficiency of cognitive radio GEO-LEO coexisting satellite network. Specifically, we maximize the sum rate of the system which can be written as
\begin{align}
R_{sum}=\sum\limits_{m=1}^M\sum\limits_{k=1}^K\sum\limits_{u=1}^{U}x_{m,u,k} (C_{m,u,k}(\boldsymbol{\eta},{\bf c})+ R_{m,u,k}(\boldsymbol{\eta}))
\end{align}
where $\boldsymbol{\eta}$ is the vector of power allocation coefficients for users at $m$ LEO beam over $k$ subcarrier such that $\boldsymbol{\eta}=[\eta_{m,0,k},\eta_{m,1,k},\eta_{m,2,k},\dots,\eta_{m,u,k},\dots,\eta_{m,U,k}]$. Accordingly, ${\bf c}=[ C_{m,1,k}, C_{m,2,k},\dots,C_{m,u,k},\dots, C_{m,U,k}]$ denotes the common data rate vector of users at $m$ LEO beam over $k$ subcarrier. The maximum sum rate can be achieved through efficient subcarrier beam assignment and power allocation at secondary LEO satellite while control the interference temperature to primary GEO satellite and guarantee the minimum data rate of LEO users. The optimization problem of sum rate maximization can be formulated as
\begin{alignat}{2}
& \underset{{(\boldsymbol{\eta},{\bf c},{\bf x},{\bf p})}}{\text{max}}\  R_{sum}\label{7}\\
s.t.&
\begin{cases}
 \mathcal C_1: \sum\limits_{m=1}^M\sum\limits_{k=1}^Kx_{m,u,k} (C_{m,u,k}+ R_{m,u,k})\geq  R_{min},\ \forall u, \\
 \mathcal C_2: \sum\limits_{u=1}^{U}x_{m,u,k} C_{m,u,k}\leq  R_{m,c,k}, \forall m,k, \\
 \mathcal C_3: f^m_{m',u',k}p_{m,k}\leq I_{th},\forall m,k,\\
 \mathcal C_4: \eta_{m,0,k}+\sum\limits_{u=1}^{U}x_{m,u,k}\eta_{m,u,k}\leq 1,\ \forall m,k, \\
 \mathcal C_5: \sum\limits_{m=1}^M\sum\limits_{k=1}^Kp_{m,k}\leq P_{tot}, \\
 \mathcal C_6: \sum\limits_{m=1}^M\sum\limits_{k=1}^K x_{m,u,k}=1, \forall u, \nonumber \\
\end{cases}
\end{alignat}
 where constraint $\mathcal C_1$ guarantees the data rate of $u$ user over $k$ subcarrier and $R_{min}$ denotes the threshold of minimum data rate. Constraint $\mathcal C_2$ ensures that the common signal can be successfully decoded by all users associated with $m$ beam over $k$ subcarrier. Constraint $\mathcal C_3$ limits the interference temperature from $m$ LEO beam to $u'$ GEO user over $k$ subcarrier. Constraint $\mathcal C_4$ control the total allocated power at each beam while constraint $\mathcal C_5$ means that the sum transmit power of all beams should not exceeds the total transmit power of LEO satellite. Then, $\mathcal C_6$ says that a user should be assigned a single beam and only one channel. In the following, we provide and discuss the proposed optimization solution.
  
\section{Proposed Solution}
It can be observed that the sum rate maximization problem in (\ref{7}) is non-convex due the rate expressions and binary variable \cite{mahmood2021optimal}. Moreover, the problem is coupled on multiple optimization variables and poses high complexity. Based on the nature of this problem, it is very challenging to obtain the Global optimal solution. Thus, we first adopt a successive convex approximation technique to reduce the complexity of the joint problem and make it more tractable \cite{7296696}. According to this technique, the original non-convex functions can be efficiently replaced by properly chosen surrogates. By applying SCA technique, the data rate of $u$ user associated with $m$ beam over $k$ subcarrier can be written as
\begin{align}
R_{m,u,k}= W \tau_{m,u,k} \log_2(\gamma_{m,u,k})+\omega_{m,u,k},
\end{align}
where $\gamma_{m,u,k}=\frac{h_{m,u,k}\eta_{m,u,k}p_{m,k}}{g^{m'}_{m,u,k}q_{m',u',k}+I_{ R_{m,u,k}}+\sigma^2}$, $\tau_{m,u,k}=\frac{\gamma_{m,u,k}}{1+\gamma_{m,u,k}}$ and $\omega_{m,u,k}=\log_2(1+\gamma_{m,u,k})-\tau_{m,u,k} \log_2(\gamma_{m,u,k})$. Similarly, we apply SCA technique for the data rate of the common message. Next we define the Lagrangian of sum rate optimization problem as
\begin{align}
    & L=-\sum\limits_{m=1}^M\sum\limits_{k=1}^K\sum\limits_{u=1}^{U}x_{m,u,k} (C_{m,u,k}(\boldsymbol{\eta},{\bf c})+ R_{m,u,k}(\boldsymbol{\eta}))+\nonumber\\&\sum\limits_{u=1}^{U}\lambda1_{n}\left( R_{min}-\sum\limits_{m=1}^M\sum\limits_{k=1}^Kx_{m,u,k} \left( C_{m,u,k}+ R_{m,u,k}\right)\right)+\nonumber\\&\sum\limits_{m=1}^M\sum\limits_{k=1}^K\lambda2_{m,k}\Bigg(\sum\limits_{u=1}^{U}x_{m,u,k} C_{m,u,k}-  R_{m,c,k}\Bigg)+\sum\limits_{m=1}^M\sum\limits_{k=1}^K\nonumber\\&\lambda3_{m,k}\Big(f^m_{m',u',k}p_{m,k}- I_{th}\Big)+\sum\limits_{m=1}^M\sum\limits_{k=1}^K\lambda4_{m,k}\Big(\eta_{m,0,k}+\sum\limits_{u=1}^{U}\nonumber\\&x_{m,u,k}\eta_{m,u,k}- 1\Big)+\lambda5\Big(\sum\limits_{m=1}^M\sum\limits_{k=1}^Kp_{m,k}- P_{tot}\Big).
\end{align}
Now applying KKT conditions \cite{khan2022energy}, and compute derivation with respect to $p_{m,k}$ as
\begin{align}
    \tau_3 p_{m,k}^3+\tau_2 p_{m,k}^2+\tau_1 p_{m,k}+\tau_0=0,\label{eq11}
\end{align}
where the values of $\tau_3$, $\tau_2$, $\tau_1$ and $\tau_0$ are defined on the top of the next page, in which $I_p=g^{m'}_{m,u,k}q_{m',u',k}$ represents the interference from GEO transmissions.
\begin{figure*}
\begin{align}
    \tau_3=&\sum\limits_{u=1}^U \sum_{j=1,j\neq u}^{U} h_{m,j,k} h_{m,u,k} \eta_{n,j,k}\eta_{m,u,k} x_{m,j,k} x_{m,u,k}(\lambda5+f_{m,k} \lambda3_{m,k}x_{m,u,k}),\nonumber
\end{align}\hrulefill 
\begin{align}
    \tau_2=& \sum\limits_{u=1}^U \sum_{j=1,j\neq u}^{U} h_{m,j,k} \eta_{m,u,k} (I_p+\sigma^2) x_{m,u,k}(\lambda5+f_{m,k} \lambda3_{m,k} x_{m,u,k})+h_{m,u,k} \eta_{m,j,k} x_{m,j,k} (\lambda5 (I_p +\sigma^2)+\nonumber\\& f_{m,k}\lambda3_{m,k}(I_p+\sigma^2)x_{m,u,k}-h_{m,j,k} \eta_{m,u,k}(\lambda2_{m,k}\gamma_{m,c,k}+\lambda1_n \gamma_{m,u,k}) W x_{m,u,k}),\nonumber
\end{align}\hrulefill 
\begin{align}
    \tau_1=& \sum\limits_{u=1}^U \sum_{j=1,j\neq u}^{U} (I_p+\sigma^2)(\lambda5 \sigma^2+f_{m,k} \lambda3_{m,k} \sigma^2 x_{m,u,k}+I_p(\lambda5+f_{m,k}\lambda3_{m,k}x_{m,u,k})+W (-h_{m,j,k} \eta_{m,u,k}(\lambda2_{m,k}\gamma_{m,c,k}+\gamma_{m,u,k}\nonumber\\&+\lambda1_u\gamma_{m,u,k}) x_{m,u,k}-h_{m,u,k}\eta_{m,j,k} x_{m,j,k}(\gamma_{m,j,k} x_{m,j,k}+\lambda2_{m,k} \gamma_{m,c,k} x_{m,u,k}+\lambda1_u \gamma_{m,u,k} x_{m,u,k}))),\nonumber
\end{align}\hrulefill 
\begin{align}
    \tau_0=& \sum\limits_{u=1}^U \sum_{j=1,j\neq u}^{U} -(I_p+\sigma^2)^2 W(\gamma_{m,j,k} x_{m,j,k}+(\lambda2\gamma_{m,c,k}+\gamma_{m,u,k}+\lambda1_u\gamma_{m,u,k})x_{m,u,k}),\nonumber
\end{align}\hrulefill 
\end{figure*}
 The solution of $p_{m,k}$ can be obtained by solving the polynomial in (\ref{eq11}) using any mathematical solver. Then, the value of $n_{m,n,k}$ can be efficiently obtained as
\begin{align}
    \eta_{m,n,k}= \frac{\mu1 \pm \sqrt{\mu2}}{\mu3},
\end{align}
where there values of $\mu1$, $\mu2$ and $\mu3$ can be written as
\begin{align}
    \mu1=&\sum_{j=1,j\neq u}^{U}-\lambda4_{m,k}(I_p+\sigma^2)+h_{m,j,k} p_{m,k} W(-\gamma_{m,j,k}\nonumber\\& x_{m,j,k}+(1+\lambda1_u)\gamma_{m,u,k}x_{m,u,k}),\nonumber
\end{align}
\begin{align}
    \mu2=&\sum_{j=1,j\neq u}^{U} 4 h_{m,j,k}(1+\lambda1_u)\lambda4_{m,k} p_{m,k} (I_p+\sigma^2) \gamma_{m,u,k} W\nonumber\\& x_{m,u,k}+(\lambda4_{m,k}(I_p+\sigma^2) h_{m,j,k} p_{m,k}(\gamma_{m,j,k}x_{m,j,k}-\nonumber\\&(1+\lambda1_u)\gamma_{m,u,k}x_{m,u,k}))^2,\nonumber
\end{align}
\begin{align}
    \mu3=&\sum_{j=1,j\neq u}^{U} 2 h_{m,j,k}\lambda4_{m,k} p_{m,k}x_{m,u,k}.\nonumber
\end{align}
Similarly, $\eta_{m,0,k}$ is computed as
\begin{align}
    \eta_{m,0,k}=\frac{\lambda2_{m,k} \gamma_{m,c,k} W}{\lambda4_{m,k}}.
\end{align}
As we can see that the considered problem is affine with respect to $ C_{m,u,k}$, we employ sub-gradient method to optimize $ C_{m,u,k}$ and the dual variables. Based on sub-gradient method, in each iteration, the value of $ C_{m,u,k}$ can be updated as
\begin{align}
 C_{m,u,k}= C_{m,u,k}+\delta (1+\lambda1_u-\lambda2_{m,k}).
\end{align}
Accordingly, the values of dual variables can be updated as
\begin{align}
&\lambda1_{n}\!=\!\lambda1_{n}\!+\!\delta \bigg( R_{min}\!-\!\sum\limits_{m=1}^M\sum\limits_{k=1}^K\!\Big(\! x_{m,u,k} C_{m,u,k}\!+\! R_{m,u,k}\!\Big)\!\!\bigg),\nonumber
\end{align}
\begin{align}
&\lambda2_{m,k}=\lambda2_{m,k}+\delta\bigg(\sum\limits_{u=1}^{U}x_{m,u,k} C_{m,u,k}- R_{m,c,k}\bigg),\nonumber
\end{align}
\begin{align}
&\lambda3_{m,k}=\lambda3_{m,k}+\delta\Big( f_{m,k}p_{m,k}- I_{th}\Big),\nonumber
\end{align}
\begin{align}
&\lambda4_{m,k}=\lambda4_{m,k}+\delta\Big(\eta_{m,0,k}+\sum\limits_{u=1}^{U}x_{m,u,k}\eta_{m,u,k}- 1\Big),\nonumber
\end{align}
\begin{align}
&\lambda5=\lambda5+\delta\Big(\sum\limits_{m=1}^M\sum\limits_{k=1}^K p_{m,k}- P_{tot}\Big).\nonumber
\end{align}
where $\delta$ is the positive step size.
\begin{algorithm}[!t]
\caption{Beam and subcarrier Allocation}
\begin{enumerate}
\item \textbf{Set} $U_m=\text{RoundUp}\Big(\dfrac{U}{M}\Big)$, $x$ = zeros(M,U,K)
\item \textbf{for} $a=1:U_m$
\item \quad\textbf{for} $b=1:M$
\item \quad\quad\textbf{for c=1:K}
\item \quad\quad\textbf{Find} $e$ such that $h_{m,e,k}=\max( h_{m,:,k})$ (where \newline\indent\quad\quad$ h_{m,:,k}$ are the channel gains of all users not \newline\indent\quad\quad assigned a beam yet)
\item \quad\quad\textbf{Set} $x_{m,u,k}=1$
\item \quad\quad\textbf{Remove} user $e$ from the list of users awaiting\newline \indent\quad\quad subcarrier and beam assignment
\item \quad\quad\textbf{end for}
\item \quad\textbf{end for}
\item \textbf{end for}
\item \textbf{Return} $x_{m,u,k}$
\end{enumerate}
\label{algo1}
\end{algorithm}

Next, we propose an efficient algorithm based on a greedy approach for subcarrier beam assignment, where users are equally distributed among all the beams in the LEO system. For $U$ LEO users in the system, each beam is transmitting data to $U_m$ subset of users where $U_m$ is computed as $U_m=\text{RoundUp}\Big(\dfrac{U}{M}\Big)$ the $\text{RoundUp}(\psi)$ function rounds up $\psi$ to the closest integer. Then each beam of every subcarrier is allocated to the user, which has maximum channel gain on the beam. After this, the assigned user is removed from the list of the users that are not allocated a beam yet. Similarly, a user is assigned to every beam. This process is repeated $U_m$ times. At the end, each beam has approximately $U_m$ users, and we get the efficient solution of $x_{m,u,k}$\footnote{Here, it is important to mention that we do not claim this technique to provide the optimal value of $x_{m,u,k}$. However, it can be seen in the results section that our algorithm provides very good performance compared to the case where channels and beams are assigned randomly.}. The detailed steps of the proposed technique are also summarized in Algorithm 1.

\section{Numerical Results and Discussion}
\begin{figure}[!t]
\centering
\includegraphics [width=0.44\textwidth]{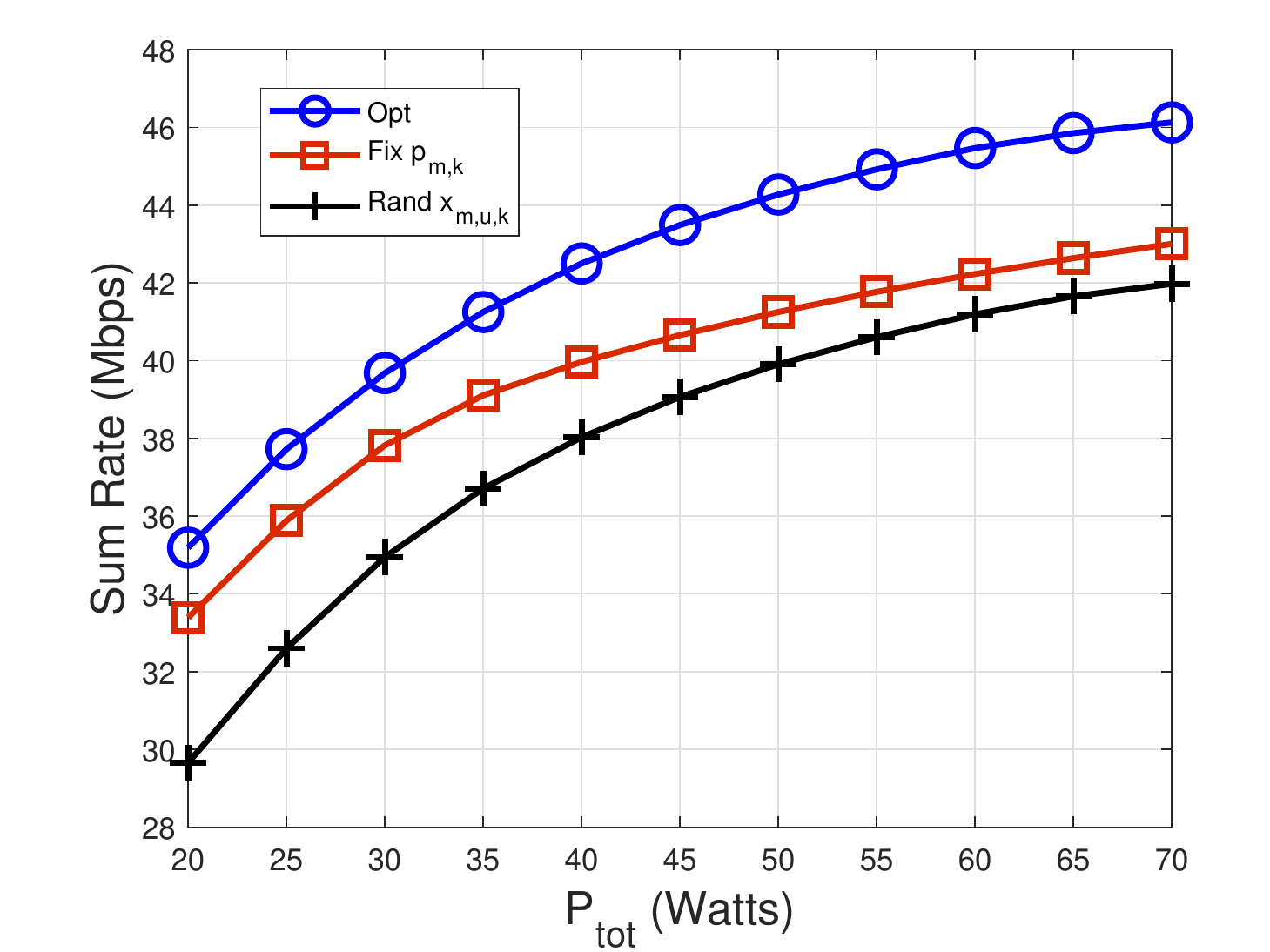}
\caption{Impact of $P_{tot}$ on the sum rate of the system and the performance comparison of different optimization frameworks.}
\label{rf2}
\end{figure}
This section provides numerical results based on Monte Carlo simulations. Unless stated otherwise we set the values of the following parameters for simulations. The number of subcarrier as $K=5$, number of beams as $M=5$, number of LEO users as $U=2K$, maximum interference threshold to GEO users as $I_{th}=2$ Watts, bandwidth of each beam as $W=10$ MHz, minimum data rate of each user as $R_{min}=1$ Mbps, total transmit power as $P_{tot}=50$ Watts, and interference from GEO satellite over each subcarrier as $I_p=4$ Watts\footnote{In the proposed framework, each beam can use a single subcarrier for communication with two RSMA users}. Moreover, we consider the frequency band as 19 GHz(Ka) and bandwidth over each beam as 10 MHz. We compare the performance of three frameworks Opt, Fix $p_{m,k}$ and Rand $x_{m,u,k}$. In Opt, we optimize the values of $x_{m,u,k}$, $p_{m,k}$, $\eta_{m,n,k}$ and $ C_{m,u,k}$ as proposed in the previous section. In fix $p_{m,k}$, the values of all other variables are optimized where as the available power is distributed equally among all the beams such that the interference threshold in not violated i.e., $p_{m,k}=min\Big(\dfrac{I_{th}}{f_{m,k}},\dfrac{P_{tot}}{MK}\Big)$. In the case of Rand $x_{m,u,k}$, all the other variables are optimized but the channel and beam are assigned to each user at random.

Figure \ref{rf2} shows the impact of increasing $P_{tot}$ on the sum rate offered by all the schemes. An increase in $P_{tot}$ results in increasing the sum rate in all proposed schemes. An interesting thing to note here is that when $P_{tot}$ is increased, the performance gap between Rand $x_{m,u,k}$ and Fix $p_{m,k}$ decreases and gap between Opt and Fix $p_{m,k}$ increases. This is because in the case of Opt and Rand $x_{m,u,k}$ when the interference threshold of a beam is met with equality, the remaining power is distributed efficiently among other beams, which can not be done in the case of Fix $p_{m,k}$. It is also clear from Fig. \ref{rf2} that the proposed Opt scheme provides the best performance for any value of $P_{tot}$. Further, it can be seen that optimizing $x_{m,u,k}$ is more beneficial than optimizing $p_{m,k}$, as Fix $p_{m,k}$ outperforms Rand $x_{m,u,k}$.
\begin{figure}[!t]
\centering
\includegraphics [width=0.44\textwidth]{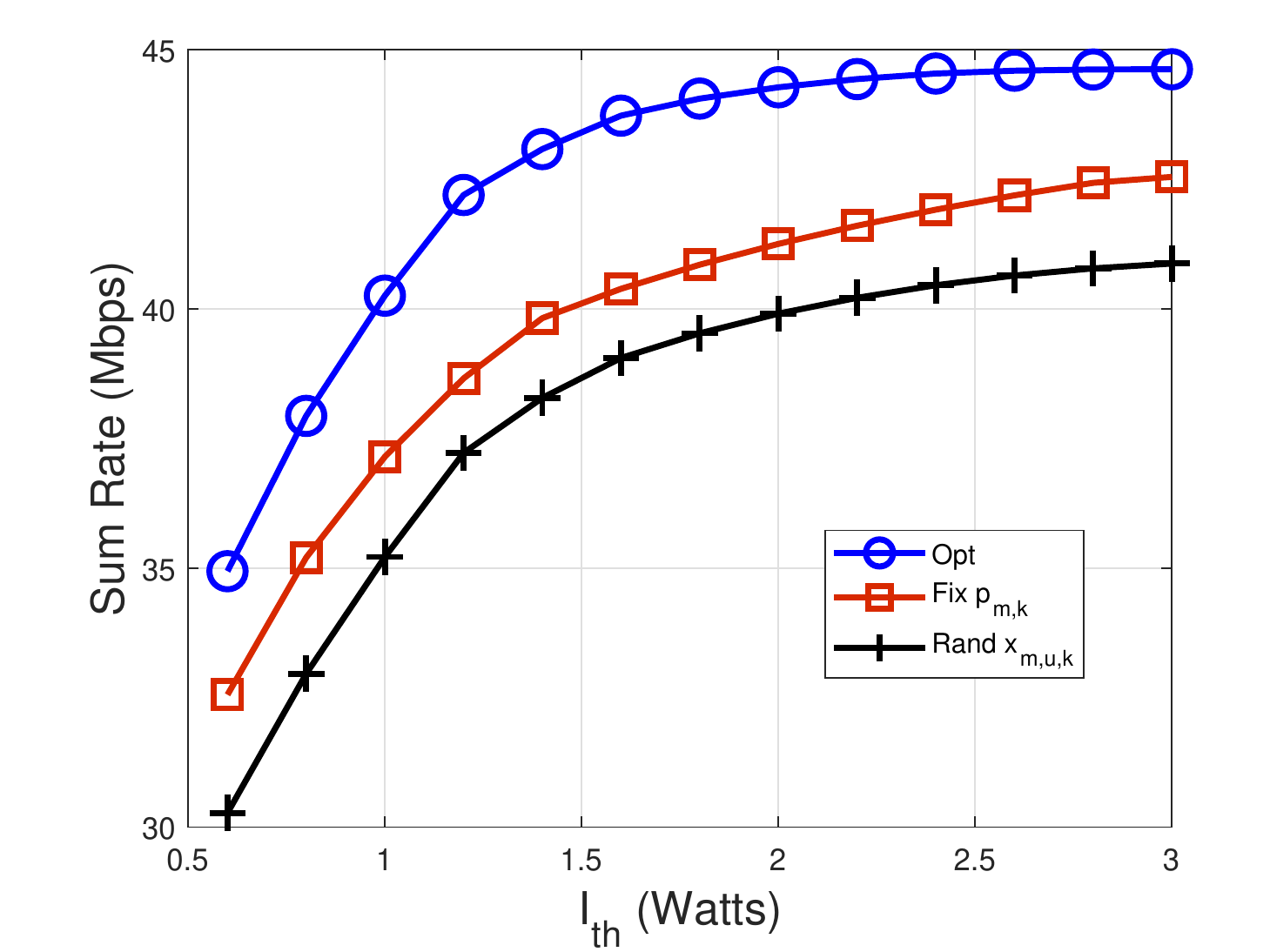}
\caption{The effect of increasing $I_{th}$ on the sum rate of the system offered by all schemes.}
\label{rf3}
\end{figure}
\begin{figure}[!t]
\centering
\includegraphics [width=0.44\textwidth]{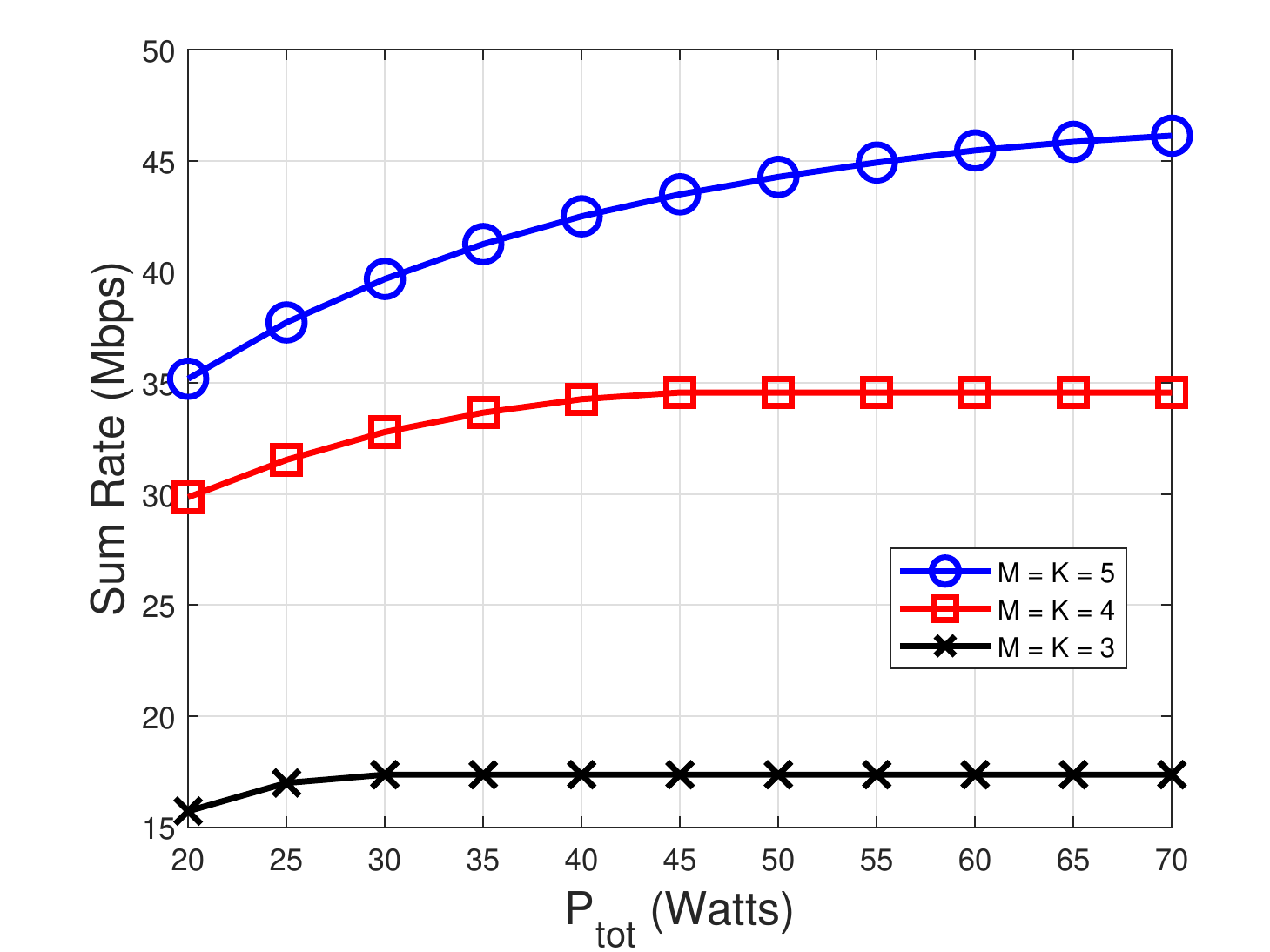}
\caption{Impact of different number of subcarrier and beams on the performance of the system.}
\label{rf4}
\end{figure}

The impact of increasing $I_{th}$ on the sum rate of the system is shown in Fig. \ref{rf3}. An increase in $I_{th}$ results in increasing the sum rate because the transmission power can be increased while satisfying the interference threshold. However, after a certain value of  $I_{th}$ when the threshold is further increased, the sum rate remains unchanged in the case of Opt scheme. Because at this point, the transmission is already being done with full available power. An interesting thing to note in Fig. \ref{rf3} is that initially, the gap between Fix $p_{m,k}$ and Rand $x_{m,u,k}$ is more, which decreases with increasing $I_{th}$ but after a certain point the gap starts to increase again. This is because at smaller values of $I_{th}$, the transmission power of all the beams is bounded by the interference threshold. When the value of $I_{th}$ is increased, the transmission power of some beams becomes unbounded by the threshold. At these points, the schemes where $p_{m,k}$ is optimized allocate the extra power from the bounded beams to other beams. Thus, the gap in performance increases. However, after a certain point, when $I_{th}$ is further increased, the transmission power of no beam is bounded by $I_{th}$. Hence, at these points, the benefit of optimizing $p_{m,k}$ increases. Therefore, the gap between Fix $p_{m,k}$ and Rand $x_{m,u,k}$ increases and performance gap between Fix $p_{m,k}$ and Opt decreases at these values of $I_{th}$. 

The impact of $K$ and $M$ on the sum rate of the system is shown in Fig. \ref{rf4}. It can be seen from the figure that the system with more $K$ and $M$ provides better performance compared to those with less $K$ and $M$. It is because more $K$ and $M$ accommodate more users which enhances the system sum rate. Further, when $P_{tot}$ is increased after a certain point there is no increase in the sum rate of the system because the transmission power of each beam ($p_{m,k}$) becomes bounded by the interference threshold ($I_{th}$). However, the point where increasing $P_{tot}$ has no impact on the sum rate comes sooner for the systems with small $K$ and $M$. 
\begin{figure}[!t]
\centering
\includegraphics [width=0.44\textwidth]{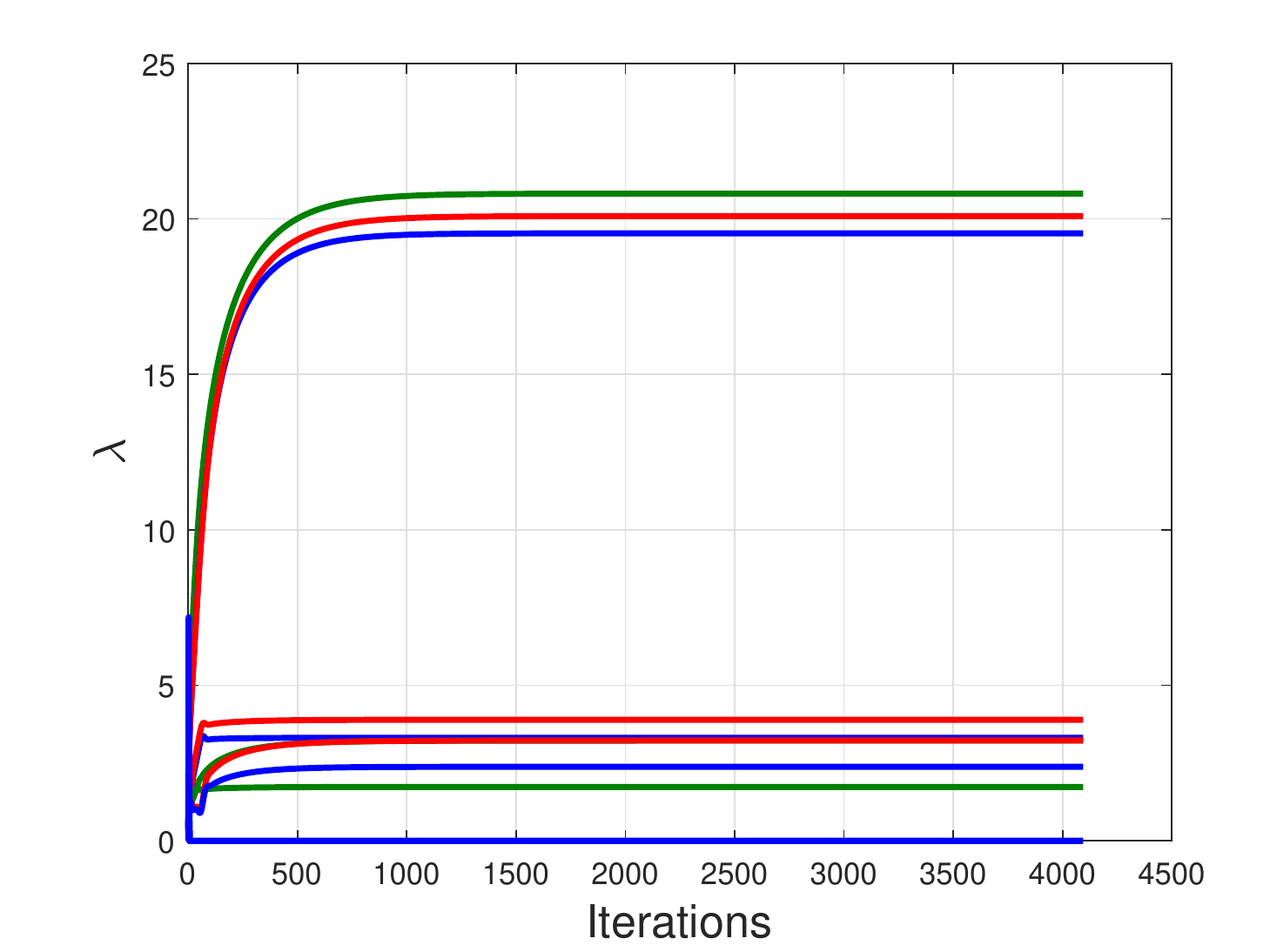}
\caption{Convergence of dual variables.}
\label{rf5}
\end{figure}

Finally, it is important to discuss the convergence complexity of the proposed scheme. Here, we show the complexity of the proposed scheme in terms of iterations required for the convergence of dual variables involved in the optimization process. The convergence behavior of the dual variables in the case of the proposed Opt scheme is shown in Fig. \ref{rf5}. It can be seen that all the dual variables converge within a reasonable number of iterations. Overall our proposed optimization framework provides significant performance in terms of the sum rate of the system by optimizing multiple variables of the system with reasonable complexity.   

\section{Conclusion}
Cognitive radio and RSMA have the potential to provide massive connectivity in space-ground communication networks. This paper has proposed RSMA for cognitive radio GEO-LEO coexisting satellite networks. Specifically, a new optimization framework for maximizing the sum rate of the secondary LEO system has been provided. The proposed framework has simultaneously optimized the transmit power of all beams, RSMA power allocation over each beam, and subcarrier beam assignment subject to each user's minimum rate and interference temperature to GEO users. Successive convex approximation technique, KKT conditions, and greedy-based algorithm have been adopted to obtain the efficient solution. The proposed optimization scheme significantly improves the system performance with reasonable complexity. This work can be further extended by jointly optimizing the resources of both LEO and GEO networks.   
\ifCLASSOPTIONcaptionsoff
  \newpage
\fi

\bibliographystyle{IEEEtran}
\bibliography{Wali_EE}

\end{document}